\documentclass[a4paper,reqno,fleqn,tbtags]{amsart}
\chardef\atcode=\catcode`\@\catcode`\@=11
\def\@settitle{\begin{flushleft}%
  \baselineskip14\p@\relax
    \bfseries
\uppercasenonmath\@title
  \@title
  \end{flushleft}}
\def\section{\@startsection{section}{1}%
  \z@{.7\linespacing\@plus\linespacing}{.5\linespacing}%
  {\normalfont\scshape}}
\def\@setauthors{%
  \begingroup
  \trivlist
 \footnotesize \@topsep30\p@\relax
  \advance\@topsep by -\baselineskip
  \item\relax
  \author@andify\authors
  {\it\authors}%
  \endtrivlist
  \endgroup}
\catcode`\@=\atcode
\usepackage[T1,T2A]{fontenc}
\usepackage[cp1251]{inputenc}
\usepackage[english,ukrainian,russian]{babel}
\usepackage{cite}
\usepackage[colorlinks]{hyperref}
\hypersetup{%bookmarks=false,
    pdfpagemode=UseNone,
    pdfstartpage=,
    pdfstartview=,
    pdfcenterwindow=true
    }
\def\N{{\inputencoding{cp1251}\selectlanguage{ukrainian}№\msp}}
\let\uu\u
\renewcommand{\P}{\mathcal P}
\newcommand{\E}{\mathcal E}
\renewcommand{\L}{\mathcal L}
\newcommand{\bP}{\boldsymbol{\mathcal P}}
\newcommand{\g}{\lVert g\rVert}
\newcommand{\Nu}{\lVert\boldsymbol u\rVert}
\renewcommand{\u}{\boldsymbol u}
\renewcommand{\U}{{\boldsymbol U\strut}}
\newcommand{\s}{\boldsymbol s}
\renewcommand{\a}{\boldsymbol a}
\renewcommand{\v}{\mathbf v}
\newcommand{\p}{\mathbf p}
\newcommand{\0}{\boldsymbol 0}
\newcommand{\2}{^{\kern.05em\boldsymbol 2}
\strut
}
\newcommand{\ld}{\,.\,}
\newcommand{\Rn}[1]{\mathrm{\uppercase\expandafter{\romannumeral#1}}}
\newbox\strutbox\setbox\strutbox=\hbox{\strut}
\newcommand\mstr{\vrule width 0pt depth 0pt height .68\ht\strutbox}
\newcommand\msp{\kern.1em}

\begin{document}
\selectlanguage{english}
\title[A Covering Second-Order Lagrangian
for the Relativistic Top]{A Covering Second-Order Lagrangian
\\for the Relativistic Top Without Forces.}
\author{Roman MATSYUK}
\address{Institute for Applied Problems in Mechanics and Mathematics, 15 Dudayev Str., L’viv, Ukraine}
\email{matsyuk@lms.lviv.ua}
\thanks{Research supported by Grants MSM:J10/98:192400002 of the Ministry of Education, Youth and
Sports, and GACR 201/00/0724 of the Grant Agency of the Czech Republic.}

\begin{abstract}
A parameter-invariant variational problem with a manifestly covariant Lagrangian function of second order is considered, which covers the case of the free relativistic top at constraint manifold of constant acceleration.
Relation to other models is discussed in brief.
\end{abstract}
\maketitle

\section{Introduction}
The interest to the description of quasi-classical physical particle by the means of higher-order
equations of motion and methods of the generalized Ostrohrads'kyj mechanics arose about
70 years ago and since then has been continuous~\cite{1,2,3,4,5,6,7}. Recently renewed attention was paid
to such models, which basically involve the notions of the first and higher curvatures of
the particle’s world line~\cite{8,9,11,12,Leiko}. In most cases, people start with an a priori given higher order
Lagrangian, and then try to interpret the dynamical system thus obtained as one describing the
motion of quasi-classical spin (the relativistic top). Technical misunderstanding of two kinds
happens to arise at this early stage. First, certain nonholonomic constraints sometimes are imposed from the very beginning. These constraints are chosen in such a way as to ensure that the Lagrangian is in fact
written in terms of the moving frame components~\cite{ 13}. But, as shown in~\cite{ 14}, nonholonomic
constraints require a more subtle approach. In particular, the constraint system does not retain
the property of variationality any more. Second, sometimes the very tempting assumption of
the unit four-velocity vector is imposed after the variation procedure has already been carried out
(cf.~\cite{15}). Such approach was quite justifiably criticized by several authors~\cite[p.\msp 149]{ 16},~\cite{ 17}. On the other hand, there exist the established systems of equations of Mathisson and Papapetrou~\cite{ 18} and of Dixon~\cite{ 19},
which are believed to be well-grounded from the point of view of physics. In 1945 Weyssenhoff~\cite{ 2}
asserted, referring to one paper of Mathisson~\cite{ 20}, `Even for a free particle in Galileian domains
the equations of motion of a material particle endowed with spin do not coincide with
the Newtonian laws of motion; there remains an additional term depending on the internal angular
momentum or spin of the particle, which raises the order of these differential equations to
three.' We add to this that the procedure of complete elimination of spin variables in fact raises
the order of the differential equations to four. In the present note this fourth order differential
equation will be shown to follow from Dixon’s form of the relativistic top equation of motion and
in case of flat space-time a Lagrange function will be proposed which produces the world lines
of thus governed spinning particle without any preliminary constraints imposed before the
variation procedure in undertaken. A constraint of constant curvature must only be imposed after the
variation procedure, and this is why we call the corresponding Lagrange function a covering Lagrangian.
\section{The relativistic top}
To start from the lowest possible order let us recall the system of Dixon equations for the quasi-classical
spinning particle in the gravitational field ($\alpha=0,1,2,3$):
\begin{equation}\label{:1}
    \dot\P^\alpha=\frac{1}{2}R^\alpha{}_{\beta\gamma\delta}u^\beta S^{\gamma\delta},\qquad\dot S^{\alpha\beta}=\P^\alpha u^\beta-\P^\beta u^\alpha.
\end{equation}
This system~(\ref{:1}) does not prescribe any preferable way of parametrization along the world line
of the particle.

It was proved in~\cite{ 21} and announced in~\cite{ 22,22a} that under the so-called auxiliary condition of
Mathisson and Pirani,
\begin{equation}\label{:2}
    u_\beta S^{\alpha\beta}=0,
\end{equation}
equations (\ref{:1}), (\ref{:2}) are equivalent to the following system of equations
\begin{subequations}
\begin{align}\label{:3}
    &\hspace*{-.3em}\varepsilon_{\alpha\beta\gamma\delta}\dot u^\beta u^\gamma u^\delta -3\,\frac{\dot u_\beta u^\beta}{\Nu^2}
    \,\varepsilon_{\alpha\beta\gamma\delta}\dot u^\beta u^\gamma s^\delta -\frac{m}{\g}\Big(\Nu^2\dot u_\alpha-\dot u_\beta u^\beta u_\alpha\Big) =\frac{\Nu^2}{2}\varepsilon_{\mu\nu\gamma\delta}R_{\alpha\beta}{}^{\mu\nu}u^\beta u^\gamma s^\delta\,,\\
    \label{:4}
    &\hspace*{-.3em}\Nu^2\dot s^\alpha+s_\beta\dot u^\beta\,u^\alpha=0,\\
    \label{:5}
    &\hspace*{-.3em}s_\alpha u^\alpha=0.
\end{align}
\end{subequations}
The correspondence between the skewsymmetric spin tensor $S^{\alpha\beta}$ and spin four-vector $s^\alpha$ under
the assumption that we recognize the Mathisson--Pirani side condition, is given by
\begin{equation*}
    s_\alpha=\frac{\g}{2\Nu}\,\varepsilon_{\alpha\beta\gamma\delta}u^\beta S^{\gamma\delta},\qquad S_{\alpha\beta}=\frac{\g}{\Nu}\,\varepsilon_{\alpha\beta\gamma\delta}u^\gamma s^\delta.
\end{equation*}
Equation~(\ref{:3}) in flat space-time was considered from variational point of view in~\cite{ 21} and some
Lagrange functions for it were proposed in~\cite{ 23}.

As promised, from now on we put $R_{\alpha\beta}{}^{\mu\nu}=0$
and proceed to eliminate the variable $s^\alpha$ from the equation~(\ref{:3}). The four-vector $s^\alpha$ in flat space-time keeps constant in all its components. To see this, one makes the contraction of the equation~(\ref{:3}) with $s^\alpha$, substitutes with it the construction $s_\beta\dot u^\beta$ in~(\ref{:4}), and takes notice of~(\ref{:5}).

In order to facilitate the calculations, it is appropriate to chose
the world line parametrization in the usual way applying the natural one: let $\U$ denote the four-vector of the unit velocity, \begin{equation}\label{:unit_velocity}
\lVert\U\rVert=1\,.
\end{equation}
and let the dots over $\U$ always denote the derivations with respect to the natural parameter.
Then we get immediately that~(\ref{:3})
takes the shape (the asterisk `$\ast$' denotes the dual tensor)
\begin{equation}\label{:6}
    \ast\,(\ddot\U\wedge\U\wedge\s)+m\,\dot\U=\0,
\end{equation}
and possesses the first integral
\begin{equation}\label{:k^2}
k^2 = \dot\U\2\equiv\dot U_\alpha\dot U^\alpha,
\end{equation}
which is nothing but the squared first curvature of the
world line.

Now we shall do some minor work on~(\ref{:6}) in order to resolve it with respect to~$\dddot\U$. First, take the dual to~(\ref{:6}) and profit from the property that $\ast^2=id$ when acting on 3-vectors in our case of $\mathrm{sign}\,(g)=-1$:
\begin{equation}\label{:6a}
    \ddot\U\wedge\U\wedge\s=-\,m\;\ast\dot\U\,.
\end{equation}
At the next step introduce technically an arbitrary vector~$\a$ and take the inner product of the equation~(\ref{:6a}) with the three-vector $\U\wedge\s\wedge\a$. On the left hand side we obtain the determinant
\begin{equation*}
    \begin{vmatrix}
       \ddot\U\cdot\U & \ddot\U\cdot\s & \ddot\U\cdot\a \\
       \U\2 & \U\cdot\s & \U\cdot\a \\
       \s\cdot\U & \s\2 & \s\cdot\a \\
     \end{vmatrix}
     =(s\2\ddot\U+k^2s\2\U)\cdot\a\,,
\end{equation*}
since $\ddot\U\cdot\U=-\,\dot\U\2=-k^2$, and $\s\cdot\U=0$ together with $\s\cdot\ddot\U=0$ in correspondence with~(\ref{:5}).

On the right hand side we notice that
\begin{equation*}
    (\ast\,\dot\U)\cdot(\U\wedge\s\wedge\a) = \frac16\varepsilon_{\alpha\beta\gamma\delta}\dot U^\alpha U^\beta s^\gamma a^\delta = \big(\ast(\dot\U\wedge\U\wedge\s)\big)\cdot\a\,,
\end{equation*}
so that finally~(\ref{:6}) becomes
\begin{equation*}
    \s\2\,(\ddot\U+k^2\U)=-m\,\ast(\dot\U\wedge\U\wedge\s).
\end{equation*}
Differentiating and then substituting the right hand side with~(\ref{:6}), we finally obtain
\begin{equation}\label{:7}
    \dddot\U+\left(k^2-\frac{m^2}{\s\2}\right)\dot\U=\0.
\end{equation}

Now let us return to the equations~(\ref{:1}) and recall the standard fact that under the Mathisson--Pirani side condition~(\ref{:2}) the particle’s momentum $\bP$ may be expressed in terms of the spin tensor $S^{\alpha\beta}$, or, equivalently,
in terms of the spin four-vector~$\s$:
\begin{equation*}
    \bP=\frac{m}{\Nu}\,\u+\frac{1}{\Nu^3}\,\ast\,\dot\u\wedge\u\wedge\s\,,
\end{equation*}
where $m = \dfrac{\bP\!\ld\u}{\Nu}$
is a constant of motion, and that the square momentum
\begin{equation*}
\bP\2=m^2-k^2\s\2+\frac{1}{\Nu^6}\big[(\dot\u\cdot\s)\,\u-(\u\cdot\s)\,\dot\u\big]\2=m^2-k^2\s\2
\end{equation*}
by virtue of~(\ref{:2}) is a constant of motion too. Thus denoting $\omega^2 = -\dfrac{\bP\2}{\s\2}$, we finally obtain the
desired fourth-order equation for the free relativistic top:
\begin{equation}\label{:8}
    \dddot\U+\omega^2\dot\U=\0.
\end{equation}

One may once more verify directly that the first Frenet curvature~$k$ of the world line as given by~(\ref{:k^2}), is the integral of~(\ref{:8}). Namely, differentiating~(\ref{:unit_velocity}) three times we obtain
\begin{equation}\label{:UdotU}
    \dot\U\U=0,\quad\dddot\U\U=-3\ddot\U\dot\U\,.
\end{equation}
Differentiating~(\ref{:k^2}), in view of~(\ref{:UdotU}), and after substituting~$\dddot\U$ with~(\ref{:8}), gives
\begin{equation}\label{:diff_k^2}
    \big(k^2\big)\boldsymbol{{}^{\dot{}}}
    =2\ddot\U\dot\U    =-\frac{2}{3}\dddot\U\U=\frac{2}{3}\omega^2\dot\U\U=0\,.
\end{equation}
\section{Variational attempt}
We shall try to find a variational equation most close to~(\ref{:8}). The strategy consists in first considering a general form of the variational equation for a parameter-invariant variational problem and then comparing it with~(\ref{:8}) in the special parametrization given by~(\ref{:unit_velocity}). Each parameter-invariant variational problem posed in coordinates $x^\alpha$ with the velocities $u^\alpha$ can be expressed in same coordinates $t=x^0$, $x^i$ ($i=1,2,3)$, but setting the velocities to
\begin{equation}\label{:v^i}
u^0 = 1,\quad v^i=\dfrac{dx^i}{dt},\quad v'^i=\dfrac{dv^i}{dt},\quad \text{etc.}
\end{equation}
The Euler--Poisson system of equations for a parameter-invariant variational problem in an arbitrary parametrization,
\begin{equation}\label{:O-P_general parametrization}
   \E_\alpha(x^\beta,u^\beta,\dot u^\beta,\ddot u^\beta,\dddot u^\beta)=0\,,
\end{equation}
by means of special coordinates~(\ref{:v^i}) takes the form

\hspace{1.9em}
\parbox{10cm}{%
$\E_\alpha=
\begin{cases}
-u^iE_i(t,x^i,v^i, v'^i, v''^i, v'''^i)=&\hspace{-8pt}0\,,
    \\
    \phantom{-}u^0E_i(t,x^i,v^i, v'^i, v''^i, v'''^i)=&\hspace{-8pt}0\,,
  \end{cases}
$
}%
    \hfill
\begin{minipage}{1cm}
    {\begin{eqnarray}
    \notag
        \\    \label{:O-P_space3component}
  \end{eqnarray}}
\end{minipage}
\endgraf
\noindent
where we express all the velocities $\v,\dotsc,\v'''$ in terms of the velocities $\u,\dotsc,\dddot\u$.
The Lagrange function for~(\ref{:O-P_general parametrization}) is given by
\begin{equation}\label{:LagrangainCal}
    \L(x^\beta,u^\beta,\dot u^\beta)=u^0L\,,
\end{equation}
where $L(t,x^i,v^i, v'^i)$ is the Lagrange function for the system~${E_i=0}$ in~(\ref{:O-P_space3component}).
If we impose the space-time homogeneity condition, then the Euler--Poisson expression in~(\ref{:O-P_space3component}) will read
\begin{equation}\label{:E_i}
    \mathbf E=\frac{d}{dt}\left(-\frac{\partial }{\partial \v}L+\frac{d}{dt}\frac{\partial }{\partial \v'}L\right).
\end{equation}

Let us investigate the question, whether there might exist an Euler--Poisson expression~$\mathbf E=(E_i)$ in~(\ref{:O-P_space3component}) generating the left hand side of~(\ref{:O-P_general parametrization}) for a parameter-invariant variational problem which, expressed in the natural parametrization by the arc length~$d\tau=\sqrt{1+\v\2}\,dt$ (the proper time), would produce the first term in~(\ref{:8}).

With that end in view we recalculate
\begin{equation}\label{:d/dt_ddot_U}
\dddot U^i\equiv\frac{d}{d\tau}\ddot U^i=\frac{dt}{d\tau} \frac{d}{dt}\ddot U^i
\end{equation}
to the time parametrization, by which
\begin{align*}
\dot t&=\frac{1}{\sqrt{1+\v\2}}\,,\\
\ddot t&=-\frac{(\v'\v)}{(1+\v\2)^2}\,,\\
\dddot t&=4\frac{(\v'\v)^2}{(1+\v\2)^{7/2}}-\frac{{\v'}\2+(\v''\v)}{(1+\v\2)^{5/2}}\,.
\end{align*}
We have
\begin{align}
U^i&=\dot t v^i\,,\notag\\
\dot U^i&=\ddot t v^i+\dot t^2 v'^i\,,\label{:dotUbyNaturalParam}\\
 \ddot U^i&=\dddot t v^i+3\,\dot t\,\ddot t v'^i+\dot t^3 v''^i\,,\notag
\\ &=\left\{4\frac{(\v'\v)^2}{(1+\v\2)^{7/2}}-\frac{{\v'}\2+(\v''\v)}{(1+\v\2)^{5/2}}\right\}v^i-3\frac{(\v'\v)}{(1+\v\2)^{5/2}}v'^i+\frac{v''^i}{(1+\v\2)^{3/2}}\,.
\notag
\end{align}

Let us put $\ddot U_i$ in the form
\begin{equation}\label{:ddot_U_2}
\ddot U_i=\left\{4\frac{(\v'\v)^2}{(1+\v\2)^{7/2}}-\frac{{\v'}\2}{(1+\v\2)^{5/2}}
-\frac{(\v''\v)}{(1+\v\2)^{5/2}}
\right\}v_i 
+\frac{d}{2dt}\frac{\partial }{\partial v'^i}\frac{{\v'}\2}{(1+\v\2)^{3/2}}.
\end{equation}
We demand that~(\ref{:d/dt_ddot_U}) coincide with~(\ref{:O-P_space3component}), that is
\begin{equation}\label{:coincides}
    \frac{d}{dt}\ddot U_i\equiv \frac{d}{dt}\left(-\frac{\partial }{\partial v^i}L+\frac{d}{dt}\frac{\partial }{\partial v'^i}L\right).
\end{equation}
First let us consider the identity
\begin{equation}\label{:ddotUcoincides}
    \ddot U_i\equiv -\frac{\partial }{\partial v^i}L+\frac{d}{dt}\frac{\partial }{\partial v'^i}L.
\end{equation}
Comparing~(\ref{:ddot_U_2}) with~(\ref{:ddotUcoincides}) it becomes obvious that for the last term in~(\ref{:ddot_U_2}) one should try
\begin{equation}\label{:L1}
    L_1=\frac{1}{2}\frac{{\v'}\2}{(1+\v\2)^{3/2}}.
\end{equation}
Substituting $L$ in~(\ref{:ddotUcoincides}) with $L_1+L_2$, where $L_1$ is given by~(\ref{:L1}), the term $\dfrac{d}{dt}\dfrac{\partial }{\partial v'^i}L_1$ on the right cancels out with the last term in~(\ref{:ddot_U_2}). After collecting like terms in~(\ref{:ddotUcoincides})
and expanding the definition of the total derivative operator, 
\begin{equation*}
 \frac{d}{dt}=v'^i\frac{\partial }{\partial v^i}+v''^i\frac{\partial }{\partial v'^i}\,,
\end{equation*}
the identity~(\ref{:ddotUcoincides}) produces the following identity on~$L_2$,
\begin{equation}\label{:a_k}
    \left\{4\frac{(\v'\v)^2}{(1+\v\2)^{7/2}}-\frac{5}{2}\frac{{\v'}\2}{(1+\v\2)^{5/2}}-\frac{(\v''\v)}{(1+\v\2)^{5/2}}\right\}\v\equiv-\frac{\partial L_2}{\partial \v}+v'^i\frac{\partial }{\partial v^i}\frac{\partial L_2}{\partial \v'}+v''^i\frac{\partial }{\partial v'^i}\frac{\partial L_2}{\partial \v'}.
\end{equation}
The coefficients of~$\v''$ suggest that one may try the expression $-\dfrac{(\v'\v)}{(1+\v\2)^{5/2}}\v$ for~$\dfrac{\partial L_2}{\partial \v'}$, and this choice integrates into 
\begin{equation}\label{:L_2}
    L_2=-\frac{(\v'\v)^2}{2(1+\v\2)^{5/2}}+L_3.
\end{equation}
Continuing the iteration, substituting $L_2$ on the right hand side of~(\ref{:a_k}) with~(\ref{:L_2}), the term $v''^i\dfrac{\partial }{\partial v'^i}\dfrac{\partial L_2}{\partial \v'}$ cancels out with the third term on the left in~(\ref{:a_k}). After collecting like terms the relationship~(\ref{:ddotUcoincides}) becomes 
\begin{equation}\label{:L_3pre_identity}
    -\frac{3}{2}\left\{\frac{{\v'}\2}{(1+\v\2)^{5/2}}-\frac{(\v'\v)^2}{(1+\v\2)^{7/2}}\right\}\v\equiv
    -\frac{\partial }{\partial\v}L_3\,. 
\end{equation}
According to~(\ref{:coincides}) one has to apply the total derivative to~(\ref{:L_3pre_identity}) and thus obtains
the following identity on~$L_3$:
\begin{equation}\label{:L_3identity}
    \frac{3}{2}\frac{d}{dt}\left\{\frac{{\v'}\2}{(1+\v\2)^{5/2}}-\frac{(\v'\v)^2}{(1+\v\2)^{7/2}}\right\}\v\equiv
    \frac{d}{dt}\frac{\partial }{\partial\v}L_3\,. \end{equation}
The expression in braces on the left hand side of~(\ref{:L_3identity}) is nothing but $\dfrac{1}{\sqrt{1+\v\2}}$ times the square of the first Frenet curvature of the world line which according to~(\ref{:k^2}) should keep to a constant value, say~${k_0}^2$, along the Euler--Poisson equation. Under this condition $L_3$ becomes a multiplier of the free relativistic particle Lagrange function,
\begin{equation}\label{:L_3}
    L_3=\frac{3}{2}{k_0}^2\sqrt{1+\v\2}.
\end{equation}
The Lagrange function for the second term in~(\ref{:8}) (set in time parametrization according to (\ref{:O-P_space3component}) and~(\ref{:dotUbyNaturalParam})) is that of free motion again and reads
\begin{equation}\label{:L_3_free}
    -\omega^2\sqrt{1+\v\2}.
\end{equation}
It remains to put together (\ref{:L1}), (\ref{:L_2}), (\ref{:L_3}) and~(\ref{:L_3_free}) to obtain
\begin{equation}\label{:completeL}
    L=\frac{1}{2}\sqrt{1+\v\2}\Bigl(k^2+3{k_0}^2-2\omega^2\Bigr),
\end{equation}
or, in terms of~(\ref{:LagrangainCal}) with $u^0\sqrt{1+\v\2}=\Nu$,
\begin{equation*}
    \L=\frac{1}{2}\Nu\Bigl(k^2+3{k_0}^2-2\omega^2\Bigr).
\end{equation*}

\section{Hamilton--Ostrohrads'kyj approach}
Let us mention one more time  that we tend to set a parameter-invariant variational problem in order to
get the world lines without any additional parametrization. Recall the general formula for the
first curvature of the world line in arbitrary parametrization,
\begin{equation}\label{:9}
    k=\frac{\lVert\dot\u\wedge\u\rVert}{\Nu^3}\,,
\end{equation}
and consider the following Lagrange function:
\begin{equation}\label{:10}
    \mathcal L=\frac{1}{2}\Nu(k^2+A).
\end{equation}
This Lagrange function constitutes a parameter-invariant variational problem because
it satisfies the Zermelo conditions:
\begin{equation}\label{:11}
    \left(\u\ld \frac{\partial}{\partial\u}+2\,\dot\u\ld\frac{\partial}{\partial\dot\u}\right)\mathcal L=\mathcal L\,,\qquad \u\ld\frac{\partial\mathcal L}{\partial\dot\u}=0\,.
\end{equation}
Variational equations are given by
\begin{equation}\label{:12}
    {}-\dot{\boldsymbol\wp}=\0\,,
\end{equation}
where
\begin{equation*}
    \boldsymbol\wp=\frac{\partial\mathcal L}{\partial\u}-\left(\frac{\partial\mathcal L}{\partial\dot\u}\right)^{\textstyle\boldsymbol{\cdot}}\;.
\end{equation*}

Now, one can calculate the quantity $\dot{\boldsymbol\wp}$ and \emph{afterwards} impose the constraint $\Nu = 1$, thus benefiting from the
parameter homogeneity of equation~(\ref{:12}). We get for~(\ref{:12}):
\begin{equation}\label{:13}
    \dddot\U+\left(\frac{3}{2}\,\dot\U\2-\frac{A}{2}\right)\dot\U+3\,(\ddot\U\cdot\dot\U)\,\U=\0.
\end{equation}

On the surface $k = k_0$ equation~(\ref{:13}) will coincide with~(\ref{:8}) if we put
\begin{equation*}
    \frac{1}{2}A=\frac{3}{2}\,k_0{}^2-\omega^2.
\end{equation*}
This completes the proof of the assertion announced in~\cite{ 24}.

To pass to the canonical formalism, it is necessary to introduce the parametrization by time, as in~(\ref{:v^i}).
In these coordinates formula~(\ref{:10}) together with~(\ref{:9}) suggest the
following expression for the Lagrange function:
\begin{equation}\label{:14}
    L=\frac{1}{2}\,\sqrt{1+\v\2}(k^2+A)\,,\qquad k^2=\frac{{\v'}\2+(\v'\times\v)\2}{\big(1+\v\2\big)^3}\,.
\end{equation}
Generalized Hamilton function $H$ should be expressed in terms of $\v$ and the couple of momenta
\begin{equation}\label{:15}
    \p^{(1)}=\frac{\partial L}{\partial\v'}\,,\qquad \p=\frac{\partial L}{\partial\v}-\frac{d}{dt}\,\p^{(1)}\,.
\end{equation}
Namely,
\begin{equation}\label{:15a}
    H=\p\ld\v+\p'\!\ld\v'-L\,.
\end{equation}

Let us proceed further to find the Hamilton function as follows. First we single out the principal terms in the Lagrange~(\ref{:14}) and Hamilton~(\ref{:15a}) functions:
\begin{equation*}
    L=L_{\Rn1}+\frac{A}{2}\sqrt{1+\v\2}\,,\qquad H=H_{\Rn1}-\frac{A}{2\,\sqrt{1+\v\2}}\,.
\end{equation*}
At the second stage we calculate the momentum~$\p_{\Rn1}{}^{(1)}$ from the first expression in~(\ref{:15}) for the Lagrange function~$L_{\Rn1}$:
\begin{equation}\label{:pI1}
\begin{split}
    \p_{\Rn1}{}^{(1)}&=\frac{\partial L_{\Rn1}}{\partial\v'}\\
                     &=\frac{\v'+\v\2\,\v'-(\v'\!\cdot\v)\,\v}{\big(1+\v\2\big)^{5/2}}\,.
\end{split}
\end{equation}
Notice, that we shall have no need in the actual expression for the momentum~$\p_{\Rn1}$ in order to find the expression for the corresponding Hamilton function. Nevertheless, as the equation of motion reads
\begin{equation*}
    {}-\frac{d\p_{\Rn1}}{dt}=\0\,,
\end{equation*}
we reproduce here the quantity~$\p_{\Rn1}$, calculated for the Lagrange function~$L_{\Rn1}$,
\begin{equation}\label{:pI}
    \begin{split}
       \p_{\Rn1} & = \frac{\partial L_{\Rn1}}{\partial\v}-\frac{d}{dt}\,\p_{\Rn1}{}^{(1)} \\
         & = -\,\frac{\v''}{\big(1+\v\2\big)^{3/2}}+3\,\frac{\v'\!\cdot\v}{\big(1+\v\2\big)^{5/2}}\,\v' +\frac{\v''\!\cdot\v}{\big(1+\v\2\big)^{5/2}}\,\v
         -\frac{{\v'}\2}{2\big(1+\v\2\big)^{5/2}}\,\v
         -\frac{5\,(\v'\!\cdot\v)^2}{2\big(1+\v\2\big)^{7/2}}\,\v\,.
     \end{split}
\end{equation}

The function~$L_{\Rn1}$ is homogeneous of order~2 in the~$\v'$ variable and due to this one immediately gets from~(\ref{:15a}) that
\begin{equation}\label{:HI}
\begin{split}
  H_{\Rn1}=\p_{\Rn1}\ld\v+L_{\Rn1}\qquad\text{with}\quad L_{\Rn1}&=\frac{k^2}{2}\,\sqrt{1+\v\2}\\
    &=\frac{{\v'}\2}{2\big(1+\v\2\big)^{3/2}}-\frac{(\v'\!\cdot\v)^2}{2\big(1+\v\2\big)^{5/2}}
\end{split}
\end{equation}

With the help of~(\ref{:pI1}) we seek to substitute the terms ${\v'}\2$ and $(\v'\!\cdot\v)^2$ in~(\ref{:HI}) with the expressions not involving the acceleration~$\v'$. The contraction of~(\ref{:pI1}) with~$\v$ and with~$\v'$ produces the following two expressions:
\begin{equation}\label{:v'v}
    \p_{\Rn1}{}^{(1)}\!\ld\v\,=\,\frac{\v'\!\ld\v}{\Big(1+\v\2\Big)^{5/2}}\,;\qquad
    \frac{{\v'}\2}{\Big(1+\v\2\Big)^{3/2}}\;=\;\p_{\Rn1}{}^{(1)}\!\ld\v'+\Big(1+\v\2\Big)^{5/2}(\p_{\Rn1}{}^{(1)}\!\ld\v)^2.
\end{equation}
With the first of the equations~(\ref{:v'v}) the expression~(\ref{:pI1}) now becomes
\begin{equation}\label{:InverseLegendre}
    \frac{\v'}{\Big(1+\v\2\Big)^{3/2}}\,=\,\p_{\Rn1}{}^{(1)}\,+\;(\p_{\Rn1}{}^{(1)}\!\ld\v)\;\v\,,
\end{equation}
which gives the lower tier of the two-level inverse of the second-order Legendre transformation~(\ref{:pI1},~\ref{:pI}).
Next we take the contraction of~(\ref{:InverseLegendre}) with~$\p_{\Rn1}{}^{(1)}$ and substitute for the~$\p_{\Rn1}{}^{(1)}\!\ld\v'$ in the second equation of~(\ref{:v'v}) to produce
\begin{equation}\label{:v'^2}
    \frac{{\v'}\2}{\Big(1+\v\2\Big)^{3/2}}=\Big(1+\v\2\Big)^{3/2}\Big[\p_{\Rn1}{}^{(1)}{}\2+
    (2+\v\2)(\p_{\Rn1}{}^{(1)}\!\ld\v)^2\Big]\,.
\end{equation}
With~(\ref{:v'^2}) together with the first of~(\ref{:v'v}), the Hamilton function~(\ref{:HI}) becomes
\begin{align*}
    H_{\Rn1}&=\p_{\Rn1}\ld\v+\frac{\Big(1+\v\2\Big)^{3/2}}{2}\left\{{\p_{\Rn1}{}^{(1)}}\2
    +\big(\p_{\Rn1}{}^{(1)}\!\ld\v\big)^2\right\}
\\
    &=\p\ld\v-\frac{A}{\vphantom{\sqrt{1+\v\2}}2}\,\frac{\v\2}{\sqrt{1+\v\2}}+\frac{\Big(1+\v\2\Big)^{3/2}}{2\vphantom{\sqrt{1+\v\2}}}\left\{{\p^{(1)}}\2
    +\big(\p^{(1)}\!\ld\v\big)^2\right\}\,,
\end{align*}
because $\p_{\Rn1}=\p-\dfrac{A}{2\vphantom{\sqrt{1+\v\2}}}\,\dfrac{\v}{\sqrt{1+\v\2}}$ along with $\p_{\Rn1}{}^{(1)}=\p^{(1)}$. The Hamilton function~$H$ now reads:
\begin{equation*}
    H=\p\ld\v+\frac{1}{2}\,\Big(1+\v\2\Big)^{3/2}\left\{{\p^{(1)}}\2+\big(\p^{(1)}\!\ld\v\big)^2\right\}-\frac{A}{2}\sqrt{1+\v\2}\,.
\end{equation*}

\section{Concluding notes}
1.~Equation~(\ref{:8}) was known to Riewe~\cite{ 15}, but its development directly from~(\ref{:1}) or from the Mathisson--Papapetrou system of equations~\cite{ 18} apparently was not obvious.\endgraf
2.~By means of the formula $k^3k_2{}^2k_3 =\Big[\U\wedge\dot\U\wedge\ddot\U\Big]$,
 which presents the relationship between
the successive curvatures of a curve in natural parametrization, we see immediately, that all
the extremals of~(\ref{:10}) have zero third curvature (cf.~\cite{ Yakupov}), and in terms of the time-like world line it means
that the particle evolves in a plane.\endgraf
3.~In~\cite{ 21} we proved by means of the generalized Ostrohrads'kyj momenta approach, that each of the successive curvatures of a curve, taken as a Lagrange function, produces the
extremals with this same curvature being the constant of motion. This was also observed by
Arod\'z with respect to only the first curvature~\cite{ 9}. But the problem of the simultaneous conservation of all the
curvatures, i.e. the variational description of helices, remains open (cf.~\cite{ 25}).\endgraf
4.~Surprisingly enough, the Lagrange function~(\ref{:10}) in fact coincides with the one, considered by
Bopp in~\cite{ 1} for the case of the motion of a charged particle in an electromagnetic field in the respect that does not include
the external four-potential itself. That the equations~(\ref{:1}) in their differential prolongation cover
both the Mathisson--Papapetrou equations of the spinning particle and the Lorentz--Dirac equations
of the self--radiating particle, was already noted in~\cite{ 23} in relation to the prediction of Barut~\cite{ 26}.
This gives still more grounds to call the expression~(\ref{:10}) the covering Lagrangian.\endgraf
5.~Following the ideas of~\cite{ 6}, we considered in~\cite{ 27} some non-local transformations which leave
invariant the exact form of the action integral
\begin{equation}\label{:16}
    \int\sqrt{\epsilon^2d\tau^2-d\alpha^2}=\int\mathcal L_\epsilon d\tau\,,
\end{equation}
where $d\alpha$ measures the rotation of the tangent to the world line during the increment $d\tau$ of the
proper time along it, so that the curvature $k$ equals $\dfrac{d\alpha}{d\tau}$. These non-local
transformations (linear in $\alpha$ and in $\tau$) were put an interpretation on them as those describing the transformation between the uniformly
accelerated frames of reference in special relativity (see~\cite[p.~18]{V.Ya.Skorobohat'ko} for a summary).

Treating in quite a formal way the variables~$\alpha$
and~$\tau$ as independent ones, one may hope that the variation of~(\ref{:16}) will produce the world
lines of constant curvature (i.e. constant relativistic acceleration). On the other hand, looking more closely
at the Lagrange function
\begin{equation}\label{:17}
    \mathcal L_\epsilon=\sqrt{\epsilon^2-k^2}\,,
\end{equation}
immediately leads to the concept of maximal acceleration $k=\epsilon$. Later this maximum acceleration has been prescribed the value
$\epsilon=\sqrt{\dfrac{c^{7\mstr}}{\hbar G}}$ (see~\cite{Scarpetta} and references therein, and also~\cite{28}).

The dynamics of a particle that propagates in the world with the `proper time'~(\ref{:16}) is given by the differentiation of the particle's velocity by that new `proper time'. This might suggest a new expression for the particle's energy~\cite{27,Scarpetta},~\cite[p. 88]{Novikov},
$$\mathcal E_\epsilon=\dfrac{m_0c^2\,dt}{\sqrt{1-k^{2\mstr}/\epsilon^{2\mstr}}\;d\tau}.$$

Two shortcomings spring up. First, the unconstrained variational problem with the Lagrange function~$\mathcal L_\epsilon\Nu$, consistently viewed as essentially a higher-order
Lagrange function, is incompatible with the idea of the constant curvature world lines.

Second, the variational problem with the Lagrange function~$\mathcal L_\epsilon\Nu$
is not parameter--invariant: the integrand $\mathcal L_\epsilon\Nu$, with $k$ given by means of~(\ref{:9}), does not satisfy
the Zermelo conditions~(\ref{:11}). The Lagrangian~(\ref{:10}) is unencumbered by these shortcomings.

\end{document}